\documentclass[12pt]{article}
\usepackage[dvips]{graphicx}
\usepackage{amsmath,amssymb}
\usepackage{pdproc}
\usepackage{epsf}
\usepackage{cite} 
\newcommand{\pd}{\partial} 
\newcommand{\beq}{\begin{eqnarray}}% can be used as {equation} or  {eqnarray}
\newcommand{\eeq}{\end{eqnarray}}

  \makeatletter 
  \def\@cite#1{[#1]} 
  \makeatother    
\begin{document}

\renewcommand{\thefootnote}{\alph{footnote}}

\title{Higgsless Electroweak Symmetry Breaking\footnote{Talk presented at 
SUSY 2004, Tsukuba, Japan, June 17-23, 2004.}}

\author{CSABA CS\'AKI}

\address{ 
Institute for High Energy Phenomenology, Newman Laboratory of Elementary
Particle Physics \\
Cornell University, Ithaca, NY 14853, USA
%%%%% You may comment out the e-mail address line below.  
\\ {\rm E-mail: csaki@lepp.cornell.edu}}

\abstract{We discuss the possibility of breaking the electroweak symmetry 
in theories with extra dimensions via boundary conditions, without a physical
Higgs scalar in the spectrum. In these models the unitarity violation 
scale can be delayed via the exchange of massive KK gauge bosons. The 
correct W/Z mass ratio can be enforced in a model in warped space with a 
left-right symmetric gauge group in the bulk. Fermion masses can be 
similarly generated via boundary conditions. In the perturbative regime
the $S$ parameter is too large in the simplest model, however this can be 
alleviated in by changing the structure of the light fermions slightly.
The major remaining issue is the inclusion of a heavy top quark without 
generating an unacceptably large shift in the $Zb\bar{b}$ coupling.}

\normalsize\baselineskip=15pt

\section{Introduction}

Extra dimensions offer new possibilities for electroweak symmetry 
breaking. Two known directions are to either identify the Higgs fields
with extra dimensional components of the gauge field ($A_{5,6}$) or
to break the symmetry via boundary conditions (BC's) and without a 
physical Higgs scalar in the spectrum. These later models have been 
dubbed Higgsless models of electroweak symmetry breaking, and will be the 
focus of this talk. There has been quite a lot of activity describing 
issues related to such models over the past 
year~\cite{CGMPT,CGPT,Nomura,BPR,CGHST,DHLR1,BN,CCGT,DHLR2,BPRS,MSU1,CCGT2,MSU2,Sekhar,Howard,Maxim,BRST,spurion,Nick,OtherHiggsless,Papucci}
and there were 
five parallel session talks on this subject at this 
conference~\cite{Nomsusy,Hoomsusy,Giacosusy,Seidelsusy,Tanabashisusy}.

First we will discuss the question of how one can possibly 
restore unitarity of the scattering of massive gauge bosons in a theory without
a Higgs scalar. Then we will show how to build a model that reproduces most
of the features of the standard model (SM) at low energies, and finally
discuss issues related to electroweak precision observables and
flavor physics.

\section{Unitarity without a Higgs}

\begin{figure}[ht]
\centerline{\includegraphics[width=0.4\hsize]{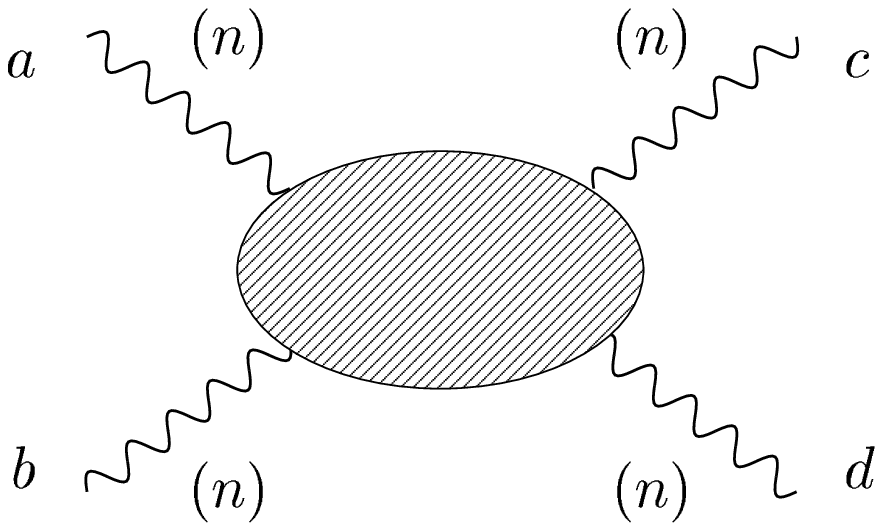}}
%\centerline{\epsfxsize=.4\textwidth\epsfbox{scattering.eps}}
\caption[]{Elastic scattering of longitudinal modes of KK gauge bosons,  $n+n\to n+n$,
with the gauge index structure $a+b\to c+d$.}
\label{fig:scattering}
\end{figure}

\begin{figure}[!ht]
\centerline{\includegraphics[width=0.75\hsize]{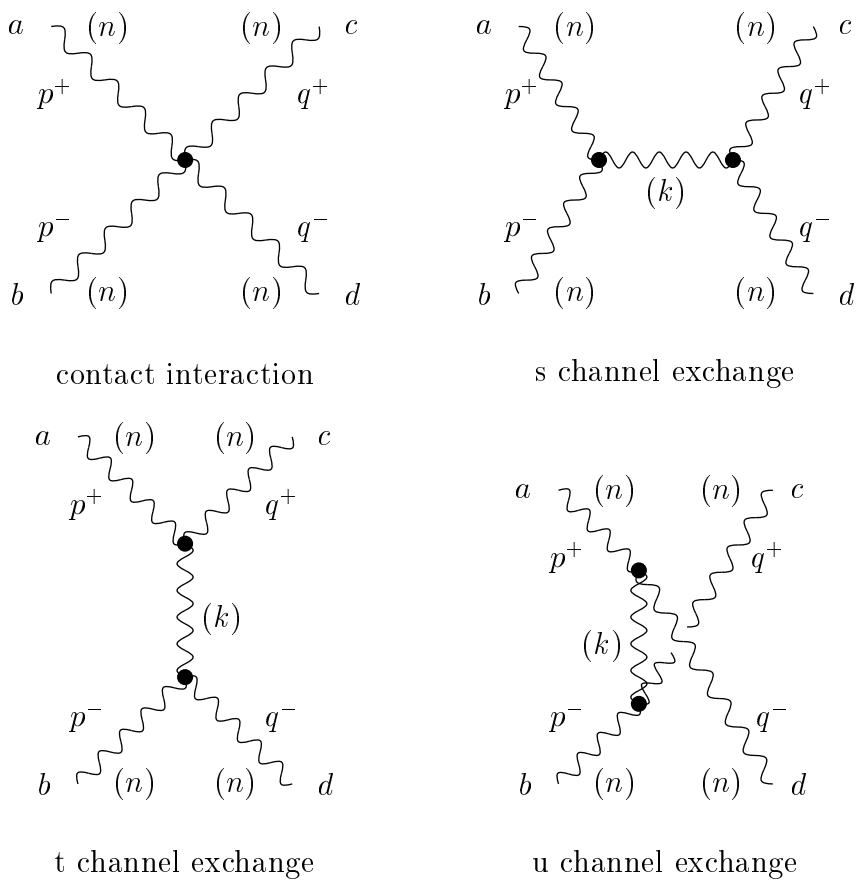}}
%\centerline{\epsfxsize=.7\textwidth\epsfbox{feynman2.eps}}
%\vspace{.7cm}
\caption{The four gauge diagrams  contributing at tree level
to the gauge boson elastic scattering amplitude.\label{fig:diagrams}}
\end{figure}

Our aim is to find a model where electroweak symmetry is broken by 
boundary conditions rather than by the expectation value of a Higgs scalar.
This immediately raises the problem of how one would ever construct a theory
that is under perturbative control beyond the 1 TeV scale. The 
reason is that scattering amplitudes of massive gauge bosons generically 
grow with energy, unless the growing terms are cancelled, which usually
happens only after the exchange of the physical Higgs scalar is added to 
the theory. In particular, the elastic scattering amplitude of the $n^{th}$
KK mode of a massive gauge field shown in Fig.\ref{fig:scattering} would
have terms that grow with the fourth and second powers of the energy. 
\begin{equation}
\mathcal{A}= A^{(4)} \frac{E^4}{M_n^4} +
A^{(2)} \frac{E^2}{M_n^2}+
A^{(0)}+{\cal O}\left(\frac{M_n^2}{E^2}\right).
 \end{equation}
In the SM $A^{(4)}$ is autumatically vanishing due to gauge invariance, 
while $A^{(2)}$ vanishes after the exchange of the physical Higgs is added. 
However, in theories with extra dimensions one needs to sum up the exchange of 
all the KK modes in order to obtain the full elastic scattering amplitude.
Since the massive KK modes themselves contain a longitudinal polarization,
there is a possibility that these scalars eaten by the massive KK modes
could play a similar unitarizing role as the Higgs does in the 
SM~\cite{CGMPT,otherunitarity}. Thus 
in order to get the full amplitude one needs to sum up the diagrams as
in Fig.~\ref{fig:diagrams}. One finds that the terms growing with energy
will be cancelled if the following two sum rules are obeyed:

\begin{eqnarray}
        \label{E4cancellation}
g^2_{nnnn} &=& \sum_k  g_{nnk}^2 ,\\
        \label{E2cancellation}
4   g_{nnnn}^2 M_{n}^2 &=& 3 \sum_k  g_{nnk}^2 M_{k}^2\, ,
\end{eqnarray}
where $g_{nnk}$ is the effective cubic coupling between the $n^{th}$, $n^{th}$
and $k^{th}$ KK mode, $ g_{nnnn}^2$ is the quartic coupling of the $n^{th}$
KK mode, and $M_n$ is the mass of the $n^{th}$ KK mode. 

We can show, that these sum rules are automatically satisfied in a 5D gauge
theory as long as 5D gauge invariance is never explicitely broken. For example
it is very simple to prove that the first sum rule is satisfied using
the completeness of the KK mode wave functions. So the problematic
terms growing with energy will always be absent in a gauge invariant 
5D theory, due to the unitarization by massive gauge bosons. Nevertheless
the theory will ultimately break down due to the linear growth with energy
of the 5D gauge coupling, and obviously the theory can only be thought of as
a low-energy effective theory valid below a cutoff scale. This 
cutoff scale can however be significantly higher than the 
unitarity violation scale of the SM without a Higgs which would be $\sim 1.7$
TeV. The actual cutoff scale of the theory is estimated from naive
dimensional analysis to be given by the scale where the one-loop factor
is approaching one due to the linear growth of the 5D coupling.
From this one obtains 
\begin{equation} \label{eq:NDA}
\Lambda_{\rm NDA} \sim \frac{24 \pi^3}{g_5^2} \, ,
\end{equation}
where $g_5$ is the dimensionful 5D coupling. 
In the Higgsless models presented below this can be expressed as 
\begin{equation}
\Lambda_{\rm NDA} \sim \frac{12 \pi^4 M_W^2}{g^2 M_{W^{(1)}}}\,,
\end{equation}
where $M_W$ is the W-mass, while $M_{W^{(1)}}$ is the mass of the 
first KK mode of the W. Thus we can see that for this scale
to be substantially above the usual unitarity violation scale of the 
SM without a Higgs which is given by $4\pi M_W/g$ one needs to have the 
first KK resonance of the W to be as light as possible. A more detailed
study of the actual scattering amplitudes including the inelastic 
channels~\cite{Papucci}
shows that the above approximation is almost correct, except that the 
cutoff scale is lower by a factor of about 4 than $\Lambda_{\rm NDA}$.

%%%%%%%%%%%%%%%%%%%%%%%%%%%%%%%%
%%%%%%%%%%%%%%%%%%%%%%%%%%%%%%%

%%%%%%%%%%%%%%%%%%%%%%%%%%%%%%%%
%%%%%%%%%%%%%%%%%%%%%%%%%%%%%%%%

\section{How to bulid a realistic Higgsless model?}

We have understood that it may be in principle possible to 
get a 5D theory without a Higgs that breaks electroweak symmetry, 
and is weakly coupled to a scale higher than the usual unitarity violation
scale of the SM would be. However, this leaves us to show that one can 
actually build a model of this sort that gives the correct W/Z mass ratio,
fermion couplings, fermion masses, etc. The first immediate question is
regarding the masses of the gauge bosons and their KK modes. In a usual
extra dimensional model the masses of the gauge fields are of the form
\begin{equation}
M_n \sim \frac{n}{R}
\end{equation}
and is usually independent of the gauge couplings. This makes it 
hard to imagine that one could actually recover the usual tree-level SM 
relation
\begin{equation}
\frac{M_W^2}{M_Z^2}=\frac{g^2}{g^2+g'^2},
\label{custodial}
\end{equation}
and leaves us also to wonder how we could ever arrange to have the KK modes
of the W and Z not appear at masses of around 2-3 times the ordinary
gauge boson masses (which would have been excluded experimentally unless 
the couplings to the SM fermions are almost exactly vanishing).

\begin{figure}[ht]
\centerline{\includegraphics[width=0.6\hsize]{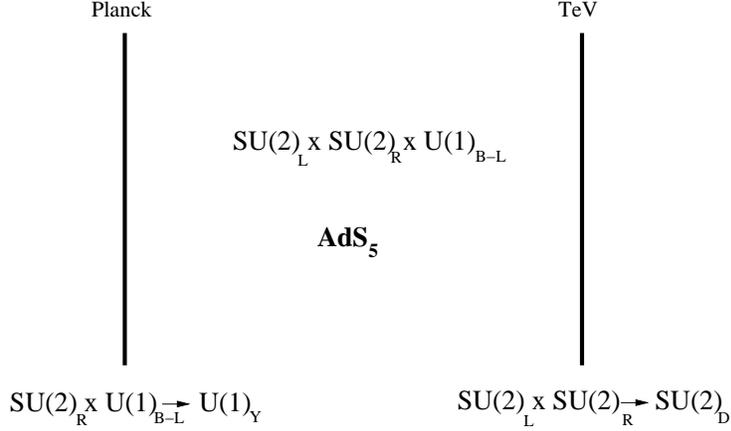}}
\caption[]{The symmetry breaking structure of the warped higgsless 
model.}
\label{fig:higgsless}
\end{figure}

The clue to finding a correct model can be obtained by understanding 
why (\ref{custodial}) is obeyed in the first place in the SM. It is 
due to the global symmetry breaking pattern in the Higgs sector. 
The Higgs potential of the SM is invariant under the rotation of all four
real components of the scalar, thus there is an approximate global 
$SO(4)\sim SU(2)_L\times SU(2)_R$ symmetry. When the Higgs gets a VEV
this group is broken to $SU(2)_D$. Thus the aim is to build an extra 
dimensional  theory where this is the actual symmetry breaking pattern,
which will the automatically imply that the relation (\ref{custodial})
has to be satisfied. One can guess~\cite{CGMPT} that there should be an 
$SU(2)_L\times SU(2)_R\times U(1)_{B-L}$ gauge group in the bulk 
of the extra dimension, but to really understand all the necessary 
parts~\cite{ADMS}
one needs to go to the AdS/CFT correspondence~\cite{holography}. 
From this one learns
that the bulk of an anti-de Sitter (AdS) 
space corresponds in ``some sense'' to a 4D conformal field theory (CFT).
Moreover, if there are some gauge fields in the bulk of the AdS space
this will imply that the CFT has a global symmetry, and the 
symmetries unbroken at the high scale (Planck brane) will remain as 
weakly gauged symmetries. While the symmetries that are broken on the 
Planck brane will be global symmetries of the CFT. Using these arguments
as a guiding principle it is clear that one needs the following setup:
there is an $SU(2)_L\times SU(2)_R\times U(1)_{B-L}$ gauge symmetry in the 
bulk of an AdS space~\cite{ADMS}, 
where the boundary conditions on the Planck brane
break $SU(2)_R\times U(1)_{B-L}$ to $U(1)_Y$, while the boundary
conditions on the TeV brane break $SU(2)_L\times SU(2)_R$ to $SU(2)_D$.
This is illustrated in Fig.~\ref{fig:higgsless}.
%%%%%%%%%%%%%%%%%%%%%%%%%%%%%%%%
%%%%%%%%%%%%%%%%%%%%%%%%%%%%%%%

Thus we will be considering a 5D gauge theory in the fixed gravitational 
background
\begin{equation}
ds^2=  \left( \frac {R}{z} \right)^2   \Big( \eta_{\mu \nu} dx^\mu dx^\nu - dz^2 \Big)
\end{equation}
where $z$ is on the
    interval $[R,R^\prime]$.  We will not be considering gravitational 
fluctuations, that we are assuming that the Planck scale is sent to infinity,
while the background is frozen to be the one given above.
In RS-type models $R$ is typically
$\sim 1/M_{Pl}$ and
    $R^\prime \sim  {\rm TeV}^{-1}$.  
The BC's corresponding to the symmetry breaking pattern discussed above is
given by~\cite{CGPT}:
\begin{eqnarray}
&
{\rm at }\  z=R^\prime:
&
\left\{
\begin{array}{l}
\partial_z (A^{L\,a}_\mu +A^{R\,a}_\mu)= 0, \
A^{L\,a}_\mu  - A^{R\,a}_\mu =0, \
\partial_z B_\mu = 0,
\\
(A^{L\,a}_5 +A^{R\,a}_5) = 0, \
\partial_z (A^{L\,a}_5 -A^{R\,a}_5)  = 0, \
B_5 = 0.
\end{array}
\right.
\label{bc1}\\
&
{\rm at }\ z=R:
&
\left\{
\begin{array}{l}
\partial_5 A^{L\,a}_\mu=0, \
    A^{R\,1,2}_\mu=0,
\\
\partial_z (g_5  B_\mu + \tilde g_5  A^{R\,3}_\mu )= 0, \
\tilde g_5 B_\mu - g_5 A^{R\,3}_\mu=0,
\\
A^{L\,a}_5=0, \ A^{R\,a}_5=0, \ B_5 = 0.
\end{array}
\right.
\label{bc2}
\end{eqnarray}
These BC's can be thought of as arising from Higgses
on each brane in the limit
of large VEVs which decouples the Higgs from gauge boson scattering
\cite{CGMPT}.
The Higgs on the TeV brane is a bi-fundamental under the two $SU(2)$'s,
while the Higgs on the Planck brane is a fundamental under $SU(2)_R$
and has charge $1/2$ under $U(1)_{B-L}$ so that
a VEV in the lower component preserves $Y=T_3+B-L$.

The linearized Maxwell equation for the bulk gauge fields in the AdS 
background is given by
\begin{equation}
(\partial_z^2 - \frac{1}{z} \partial_z  +q^2) \psi(z)=0,
\end{equation}
where the solutions in the bulk are assumed to be of the form
$A_\mu (q) e^{-iqx} \psi(z)$. The the KK mode expansion is given by
the solutions to this equation which are
of the form
\begin{equation}
	\label{eq:Bwv}
\psi^{(A)}_k(z)=z\left(a^{(A)}_k J_1(q_k z)+b^{(A)}_k Y_1(q_k z)\right)~,
\end{equation}
where $A$ labels the corresponding gauge boson.
Due to the mixing of the various gauge groups, the KK decomposition is
slightly complicated but it is obtained
by simply enforcing the BC's:
\begin{eqnarray}
           \label{eq:KKB}
B_\mu (x,z) & = & g_5\,  a_0 \gamma_\mu (x)
+   \sum_{k=1}^{\infty} \psi^{(B)}_k (z)  \, Z^{(k)}_\mu (x)\, ,
\\
           \label{eq:KKAL3}
A^{L\, 3}_\mu (x,z) & = &
{\tilde g_5} \, a_0 \gamma_\mu (x)
+   \sum_{k=1}^{\infty} \psi^{(L3)}_k (z) \, Z^{(k)}_\mu (x) \, ,
\\
           \label{eq:KKAR3}
A^{R\, 3}_\mu (x,z) & = &
{\tilde g_5}  \, a_0 \gamma_\mu (x)
+    \sum_{k=1}^{\infty}  \psi^{(R3)}_k (z) \, Z^{(k)}_\mu (x) \, ,
\\
           \label{eq:KKALpm}
A^{L\, \pm}_\mu (x,z) & = &
  \sum_{k=1}^{\infty}  \psi^{(L\pm)}_k (z) \, W^{(k)\, \pm}_\mu (x) \, ,
\\
           \label{eq:KKARpm}
A^{R\, \pm}_\mu (x,z) & = &
  \sum_{k=1}^{\infty} \psi^{(R\pm)}_k  (z) \, W^{(k)\, \pm}_\mu (x)  \, .
\end{eqnarray}
Here $\gamma(x)$ is the 4D photon, which has a flat wavefunction due to
the
unbroken $U(1)_Q$  symmetry, and $W^{(k)\, \pm}_\mu (x)$ and
$Z^{(k)}_\mu (x)$ are the KK towers of the massive $W$ and $Z$
gauge bosons, the lowest of which are supposed to correspond to the
observed $W$ and $Z$.
To leading order in $1/R$ and for
$\log \left(R^\prime/R \right) \gg1$, the lightest solution for this
equation for the mass of the $W^\pm$'s is
\begin{equation}
M_W^2 = \frac{1}{R^{\prime 2} \log \left(\frac{R^\prime}{R}\right)} \, .
\end{equation}
Note, that this result  does not depend
on the 5D gauge coupling, but only on the scales $R,R'$. 
Taking $R= 10^{-19}$ GeV$^{-1}$ will fix $R^\prime= 2 \cdot 10^{-3}$ 
GeV$^{-1}$. The lowest mass of the $Z$ tower is approximately given by
\begin{equation}
M_Z^2  = \frac{g_5^2+2 \tilde g_5^{2}}{g_5^2+ \tilde g_5^{2}}
\frac{1}{R^{\prime 2} \log \left(\frac{R^\prime}{R}\right)}  \, .
\end{equation}
If the SM fermions are localized on the Planck brane then the leading order
expression for the 4D Weinberg angle will be given by
\begin{equation}
\sin \theta_W = \frac{ \tilde g_5}{\sqrt{ g_5^2+2 \tilde g_5^2} }=
\frac{ g^\prime}{\sqrt{ g^2+  g^{\prime 2}} }.
\label{SM3}
\end{equation}
Thus we can see that to leading order the SM expression for the W/Z mass ratio
is reproduced in this theory as expected. 
In fact the full structure of the SM coupling is 
reproduced at the leading order in $1/\log R'/R$, which implies 
that at the leading log level there is no 
$S$-parameter either. An $S$-parameter in this language would have manifested
itself in an overall shift of the coupling of the Z compared to its
SM value evaluated from the $W$ and $\gamma$ couplings, which 
are absent at this order of approximation.
The corrections to the SM relations will appear in the next order 
of the log expansion, for which we will be showing the results later on.
To evaluate the predictions of this model to a precision required by the 
measurements of the electroweak observables one needs to calculate at least 
the next order of corrections to the masses and couplings, 
together with the loop effects of the 
KK gauge bosons, and subtract the usual Higgs contributions.

\section{Fermion masses}
In the SM the Higgs serves a double purpose: it gives mass to the 
gauge bosons by breaking the electroweak symmetry, but it also 
provides a mass to the fermions via the Yukawa couplings. So our next
task is to show that in the above presented warped higgsless model one
can also reproduce the fermion masses via boundary 
conditions~\cite{CGPT,Nomura,BPR,CGHST}. 
For this we need to first decide where we put the SM fermions in this
theory. If we were to put them on the TeV brane then the fermions would 
only transform under $SU(2)_D\times U(1)_{B-L}$ which would imply that the
spectrum is non-chiral. If the fermions were totally localized on the 
Planck brane then one would not be able to generate masses for them. So
the only possible way is if the fermions are in the bulk, and thus feel
both the Planck brane and the TeV brane. However, 5D fermions are Dirac
fermions, thus for every 4D Weyl fermion one needs to introduce a 
Dirac fermion. The boundary conditions will then  be chosen such that 
(as usually in a KK theory) the zero mode spectrum will be chiral.
Next we summarize briefly the properties of 5D fermions in warped space.
The action of a 5D fermion in warped space is generically given by
\begin{equation}
        \label{eq:curvedaction}
S = \int d^5 x 
\sqrt{g}
\left( 
\frac{i}{2} (\bar{\Psi}\, e_a^M \gamma^a D_M \Psi 
- D_M \bar{\Psi} \, e_a^M \gamma^a \Psi)
-m \bar{\Psi} \Psi
\right),
\end{equation}
where $e_a^M$ is the generalization of the vierbein to higher dimensions (``f\"unfbein'') satisfying
\begin{equation}
e_a^M \eta^{ab} e^N_b = g^{MN},
\end{equation}
the $\gamma^a$'s are the usual Dirac matrices, and $D_M$ is the covariant derivative including the 
spin connection term.
For the AdS$_5$ metric in the conformal coordinates written above, 
$e_M^a =(R/z) \delta_M^a$, and
$D_\mu \Psi =( \partial_\mu+\gamma_\mu\gamma_5/(4z))\Psi$, $D_5 \Psi=\partial_5 \Psi$, 
however the spin connection terms involved in the two covariant derivatives of (\ref{eq:curvedaction})
cancel each other and thus do not contribute in total to the action. 
Finally, in terms of two component spinors, the action is  given by
\begin{equation} 
S = \int d^5 x 
\left(\frac{R}{z}\right)^4 
 \left( 
- i \bar{\chi}  \bar{\sigma}^\mu \partial_\mu \chi 
- i \psi  \sigma^{\mu} \partial_\mu \bar{\psi} 
+ \frac{1}{2} ( \psi \overleftrightarrow{\partial_5} \chi 
-  \bar{\chi}  \overleftrightarrow{\pd_5} \bar{\psi} )
+ \frac{c}{z} \left( \psi \chi + \bar{\chi} \bar{\psi} \right) 
\right), 
\end{equation} 
where $c$ is the bulk Dirac mass in units of the AdS curvature $1/R$, and again $\overleftrightarrow{\partial_5}  = \overrightarrow{\partial_5}-\overleftarrow{\partial_5}$
with the convention that the differential operators act only on the spinors and not on the metric factors.
The localization properties of the zero modes are strongly dependent on the
parameter $c$.
For $c>1/2$ the left handed zero mode
is localized near the Planck brane. Conversely, for $c<1/2$
the zero mode is thus localized near the TeV brane.
In the AdS/CFT language~\cite{holography} this corresponds to the fact that for $c>1/2$ the fermions will be elementary
(since they are localized on the Planck brane), while for $c<1/2$ they are to be considered as composite
bound states  of the  CFT modes (since they are peaked on the TeV brane). 
A right handed zero mode is localized on the Planck brane for $c<-1/2$, and
on the TeV brane for $c>-1/2$. 

We now discuss how to obtain the masses for the lepton sector of the 
warped higgsless model. The quark sector can be obtained similarly with 
slight modifications. 

As always in a left--right symmetric model,  the fermions are in the representations 
$(2,1,-1/2)$ and $(1,2,-1/2)$ of SU(2)$_L\times$ SU(2)$_R\times$ U(1)$_{B-L}$ for left and right handed 
leptons respectively. Since we assume that the fermions live in the bulk, both of these 
are Dirac fermions, thus every chiral SM fermion is doubled (and the right handed neutrino is added 
similarly). Thus the left handed doublet can be written as 
\begin{equation} 
        \label{leftlep}
\left( 
\chi_{\nu_L},
\bar{\psi}_{\nu_L},
\chi_{e_L},
\bar{\psi}_{e_L}
\right)^t
\, ,
\end{equation} 
where $(\chi_{\nu_L},\chi_{e_L})$ will eventually correspond to the SM SU(2)$_L$ doublet
and $({\psi}_{\nu_L},{\psi}_{e_L})$ is its SU(2)$_L$ antidoublet partner needed to form a complete
5D Dirac spinor. Similarly, the content of the right handed doublet is 
\begin{equation} 
\label{rightlep}
\left(
\chi_{\nu_R}, \bar{\psi}_{\nu_R}, \chi_{e_R}, \bar{\psi}_{e_R}
\right)^t\, ,
\end{equation} 
where $(\psi_{\nu_R},\psi_{e_R})$ would correspond to the 'SM' right-handed doublet, i.e., the right
electron and the extra right neutrino, while $(\chi_{\nu_R},\chi_{e_R})$ is its antidoublet partner
again needed to form a complete 5D Dirac spinor.
Without boundary terms the BC's would be just $\chi_R=\psi_L=0$ on both 
branes. This will ensure that the zero modes will be given by 
$\chi_L$ and $\psi_R$. 
In order to make the zero modes acquire small masses, one can add a Dirac mass
on the TeV brane. This is because on the TeV brane the theory is non-chiral.
This will modify the BC's on the TeV brane to
\begin{eqnarray} 
&& 
{\psi_L}_{|R^{\prime\,-}}
=
-M_D R^\prime \, {\psi_R}_{|R^{\prime\, -}} 
        \label{TeV} \\ 
&& 
{\chi_R}_{|R^{\prime\, -}} 
=  
M_D R^\prime \, {\chi_L}_{|R^{\prime\, -}} 
 \end{eqnarray}
On the Planck brane the unbroken gauge group is $SU(2)_L\times U(1)_Y$ so
we can add a Majorana mass to the right handed neutrino on the Planck 
brane. This will lead to a BC of the form
\begin{equation}
\psi_{\nu_R} = MR\, \chi_{\nu_R}.
\end{equation}

Together with the bulk equations of motion these BC's lead
to the following approximate mass spectrum for the neutrinos
\begin{equation}
m_0 \sim \frac{f M_D^2 }{M^2},
\end{equation}
which is of the typical see-saw type since the Dirac mass, $M_D$, which is of the same order as that of the  electron mass, is suppressed by the large masses of the right handed neutrinos 
localized on the brane. 
Similarly one can show that a realistic spectrum is achievable in this simple 
toy model for the charged leptons is also achievable. 

\section{Electroweak precision observables}

Next we will discuss the leading corrections to the electroweak precision
observables in the warped higgsless model~\cite{BPR,DHLR1,BN,CCGT,CCGT2}.
In the following we will use the oblique corrections
$S$, $T$ and $U$ to fit the
$Z$-pole observables, mainly measured at LEP1.
These three parameters are sufficient for predicting all of those
observables.
In \cite{BPRS}, Barbieri {\it et al.} proposed a new enlarged set of
parameters to also take into account the differential
cross section measurements at LEP2. The only additional information
contained in these parameters is the bound on the
coefficients of the four-Fermi operators that are generated by the
exchange of gauge boson KK modes.In our approach we simply use the bounds on $S,T$ and $U$ from the
$Z$-pole observables, while the bounds on the four Fermi
operators are taken into account by directly imposing the constraints
on new gauge bosons from LEP2 and from the direct searches at Tevatron.

Perturbatively the $S$ parameter ``counts" the number of degrees of
freedom that participate in
the electroweak sector, while the $T$ parameter measures the amount of
additional isospin breaking. Contributions to $U$ are typically very
small. Both $S$ and $T$
must be typically small
($< 0.25$) in order
to be compatible with precision electroweak measurements.

\begin{figure}[tb]
\begin{center}
\includegraphics[width=11cm]{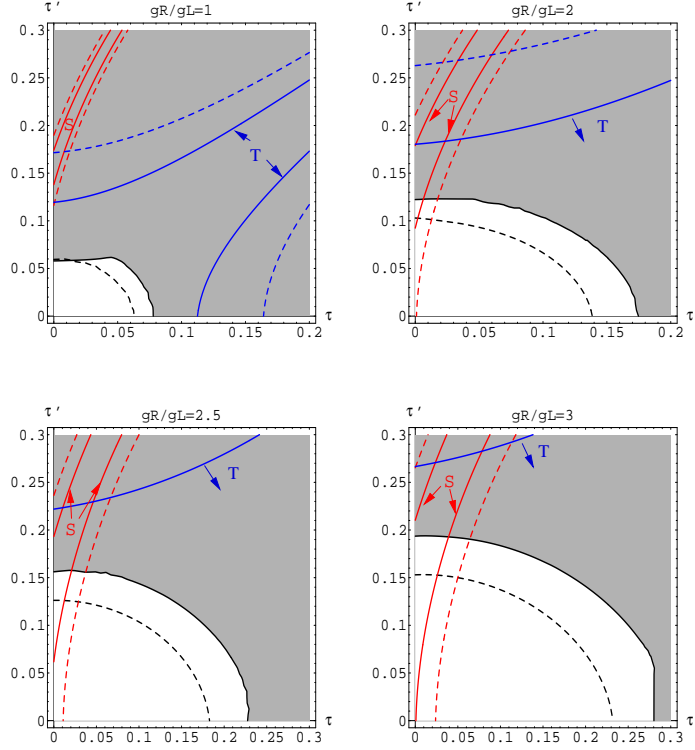}
\end{center}
\caption{Combined plots of the experimental constraints on Higgsless
models for different values of the $g_{5R}/g_{5L}$ ratio, in the
parameter space $\tau$--$\tau'$ (normalized by $R \log R'/R$).
The solid contours for $S$ (red) and $T$ (blue) are at $0.25$; the
dashed contours at $0.5$.
The black solid (dashed) line corresponds to a deviation in the
differential cross section of $3\%$ ($2\%$) at LEP2.
The shaded region is excluded by a deviation larger that $3\%$ at LEP
and/or direct search at Run1 at Tevatron.} \label{fig:combo}
\end{figure}

Electroweak symmetry breaking sectors that are more complicated than
a 4D Higgs doublet tend to have positive $S$ parameters of order 1.
In Higgsless models with
a warped extra dimension it has been shown \cite{CCGT} that both the
ratio of $SU(2)_L$ and $SU(2)_R$ couplings, $g_{5R}/g_{5L}$ as well as
kinetic terms on the TeV brane affect the $S$ and $T$ parameters
in important ways. With no brane kinetic terms and $g_{5R}=g_{5L}$,
$S=1.15$
and $T=0$. Increasing the ratio $g_{5R}/g_{5L}$ reduces $S$ to
\beq
S \approx \frac{6 \pi}{g^2 \log \frac{R^\prime}{R}}
\frac{2}{1+\frac{g_{5R}^2}{g_{5L}^2}}
\eeq
while keeping $T\approx 0$.
A qualitatively similar effect is induced by Planck brane kinetic
terms whose dimensionless coefficient we denote by $r$, 
the only difference being in the couplings of the gauge bosons,
thus affecting the bounds on direct $Z'$ searches.
It was shown in~\cite{CCGT} that the TeV brane kinetic terms
produce further corrections. We denote the dimensionless coefficient
of the $SU(2)_D$ kinetic term on the TeV brane $\tau$ and for the
$U(1)_{B-L}$ by $\tau'$. The non-Abelian
brane kinetic term gives a correction to $S$ at first order,
multiplying the previous result by $1+ \frac{4}{3} \frac{\tau}{R}$,
while giving
a very small positive contribution to $T$.
The $\tau^\prime$ corrections are more complicated, and more
interesting.  The first effects appear
at quadratic order, and they give negative corrections to both $S$ and
$T$. The Abelian
brane kinetic term, $\tau^\prime$, also has the effect of reducing the
mass of the lightest neutral
KK gauge boson resonance.
We scanned the model in this 3D parameter space,
$(g_{5R}/g_{5L},\tau,\tau^\prime)$, to uncover regions allowed by
experiments.
In Fig.\ref{fig:combo} we show combined plots for four values of
$g_{5R}/g_{5L}$= 1, 2, 2.5, 3.
In order to satisfy both precision tests and LEP2/Tevatron bounds, a
large $g_{5R}/g_{5L}$ ratio is required.
In this case, however, the masses of the resonances are raised,
making them possibly ineffective in restoring partial wave unitarity
and leading to strong coupling below $2$~TeV.
These results are in agreement with the conclusions of
refs.~\cite{BPRS,DHLR2}.

In the following, we would like to focus on an
alternative solution~\cite{CCGT2} to the $S$  problem which has additional
beneficial side-effects.
It has been known for a long time in Randall-Sundrum (RS) models with a
Higgs that the
effective $S$ parameter
is large and negative \cite{CET}  if the fermions are localized on the
TeV brane as
originally proposed. When the fermions
are localized on the Planck brane the contribution to $S$ is positive,
and so for some intermediate localization the $S$ parameter vanishes,
as first pointed out for RS models by Agashe et al.\cite{ADMS}. The
reason for this is fairly simple.  Since the $W$ and $Z$ wavefunctions
are approximately flat, and the gauge KK mode wavefunctions are
orthogonal to them, when the fermion wavefunctions are
also approximately flat the overlap of a gauge KK mode with two
fermions will approximately vanish. Since it is the coupling of the
gauge KK modes to the fermions that induces a shift in the $S$
parameter, for approximately flat fermion wavefunctions the $S$
parameter must be small.
Note that not only does reducing the coupling to gauge KK modes reduce
the $S$ parameter, it also weakens the experimental constraints on the
existence of light KK modes.
This case of delocalized bulk fermions is not covered by the no--go
theorem of~\cite{BPRS}, since there it was assumed that the fermions
are localized on the Planck brane.

In order to quantify these statements, it is sufficient to
consider a toy model where all the three families of fermions are
massless and have a universal delocalized profile in the bulk.
Before showing some numerical results, it is useful to understand the
analytical behavior of $S$ in interesting limits.
For fermions almost localized on the Planck brane, it is possible to
expand the result for the $S$-parameter in powers of $(R/R')^{2c_L-1} \ll 1$.
The leading terms, also expanding in powers of $1/\log$, are:

\beq
S = \frac{6 \pi}{g^2\, \log \frac{R'}{R}} \left( 1 - \frac{4}{3}
\frac{2c_L-1}{3-2c_L} \left( \frac{R}{R'}\right)^{2c_L-1} \log
\frac{R'}{R} \right)~,
\eeq \label{eq:Splanck}
and $U \approx T \approx 0$.
The above formula is actually valid for $1/2 < c_L < 3/2$. For
$c_L>3/2$ the corrections are of order $(R'/R)^2$ and numerically
negligible.
As we can see, as soon as the fermion wave function starts leaking into
the bulk, $S$ decreases.

Another interesting limit is when the profile is almost flat, $c_L
\approx 1/2$.
In this case, the leading contributions to $S$ are:

\beq
S =  \frac{2 \pi}{g^2\, \log \frac{R'}{R}} \left( 1 + (2 c_L -1)\, \log
\frac{R'}{R} + \mathcal{O} \left((2c_L-1)^2 \right)\right)~.
\eeq
In the flat limit $c_L=1/2$, $S$ is already suppressed by a factor of 3
with respect to the Planck brane localization case.
Moreover, the leading terms cancel out for:
\beq
c_L = \frac{1}{2} - \frac{1}{2\, \log \frac{R'}{R}} \approx 0.487~.
\eeq

For $c_L<1/2$, $S$ becomes large and negative and, in the limit of TeV
brane localized fermions ($c_L \ll 1/2$):

\beq
S =  - \frac{16 \pi}{g^2} \frac{1-2 c_L}{5-2 c_L}~,
\eeq
while, in the limit $c_L\rightarrow - \infty$:
\begin{eqnarray}
T&\rightarrow& \frac{2 \pi}{g^2\, \log \frac{R'}{R}} (1 + \tan^2
\theta_W) \approx 0.5~,\\
U&\rightarrow& - \frac{8 \pi}{g^2\, \log \frac{R'}{R}} \frac{\tan^2
\theta_W}{2 + \tan^2 \theta_W} \frac{1}{c_L} \approx 0~.
\end{eqnarray}

\begin{figure}[tb]
\begin{center}
\includegraphics[width=12cm]{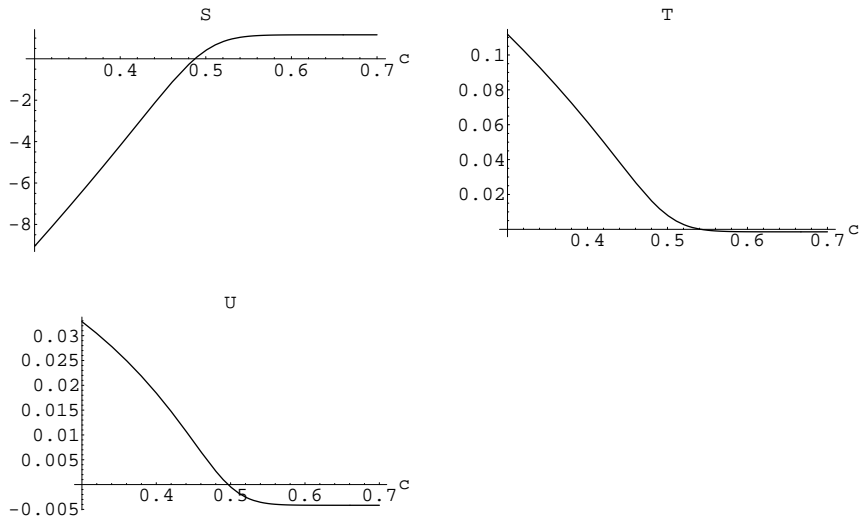}
\end{center}
\caption{
Plots of the oblique parameters as function of the bulk mass of the
reference fermion. The values on the right correspond to localization
on the Planck brane. $S$ vanishes for $c=0.487$.
} \label{fig:STUvsC}
\end{figure}

In Fig.~\ref{fig:STUvsC} we show the numerical results for the oblique
parameters as function of $c_L$.
We can see that, after vanishing for $c_L \approx 1/2$, $S$ becomes
negative and large, while $T$ and $U$ remain smaller.
With $R$ chosen to be the inverse Planck scale, the first KK resonance
appears around $1.2$~TeV, but for larger values of $R$ this scale can
be safely reduced down below a TeV.
As already discussed in the previous section, such resonances will be
weakly coupled to almost flat fermions and can easily avoid the strong
bounds from direct searches at LEP or Tevatron.
If we are imagining that the AdS space is a dual description
of an approximate conformal field theory (CFT), then $1/R$ is the scale
where the CFT is no longer
approximately conformal and perhaps becomes asymptotically free. Thus it
is quite reasonable that the scale $1/R$ would be much smaller than the
Planck scale.

\begin{figure}[tb]
\begin{center}
\includegraphics[width=10cm]{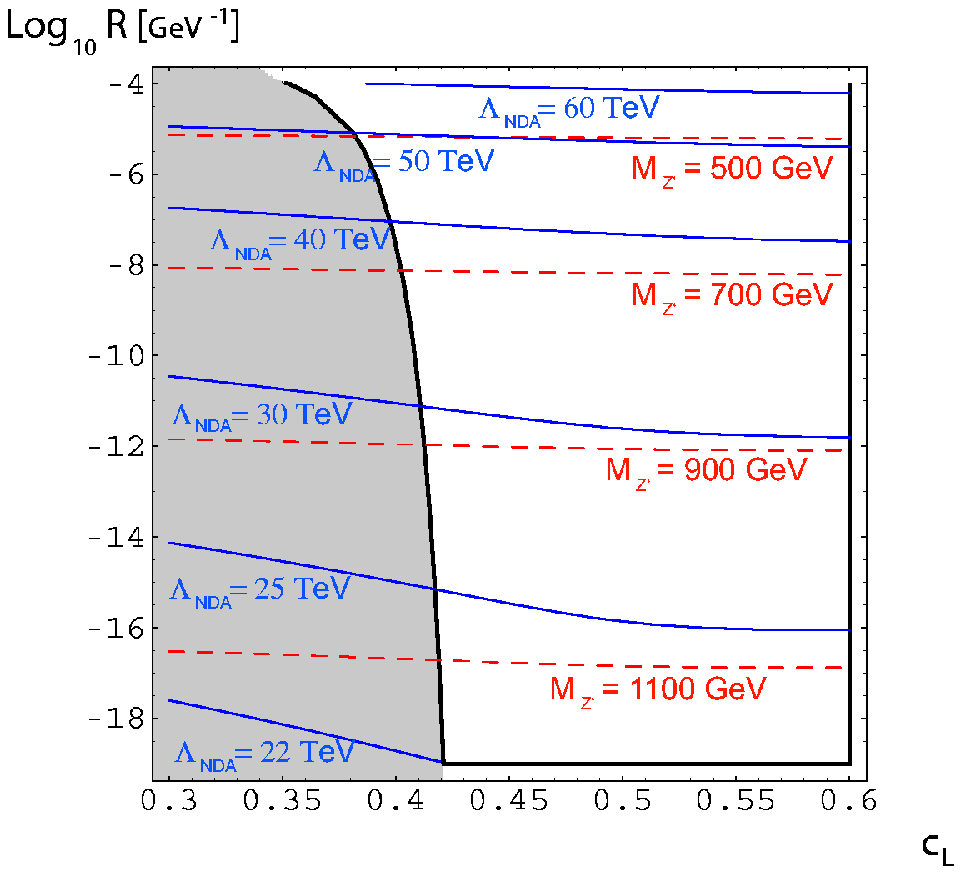}
\end{center}
\caption{Contour plots of $\Lambda_{\rm NDA}$ (solid blue lines) and
$M_{Z^{(1)}}$ (dashed red lines) in the parameter space $c_L$--$R$.
The shaded region is excluded by direct searches of light $Z^\prime$ at
LEP.
}
\label{fig:NDA}
\end{figure}

\begin{figure}[tb]
\begin{center}
\includegraphics[width=14cm]{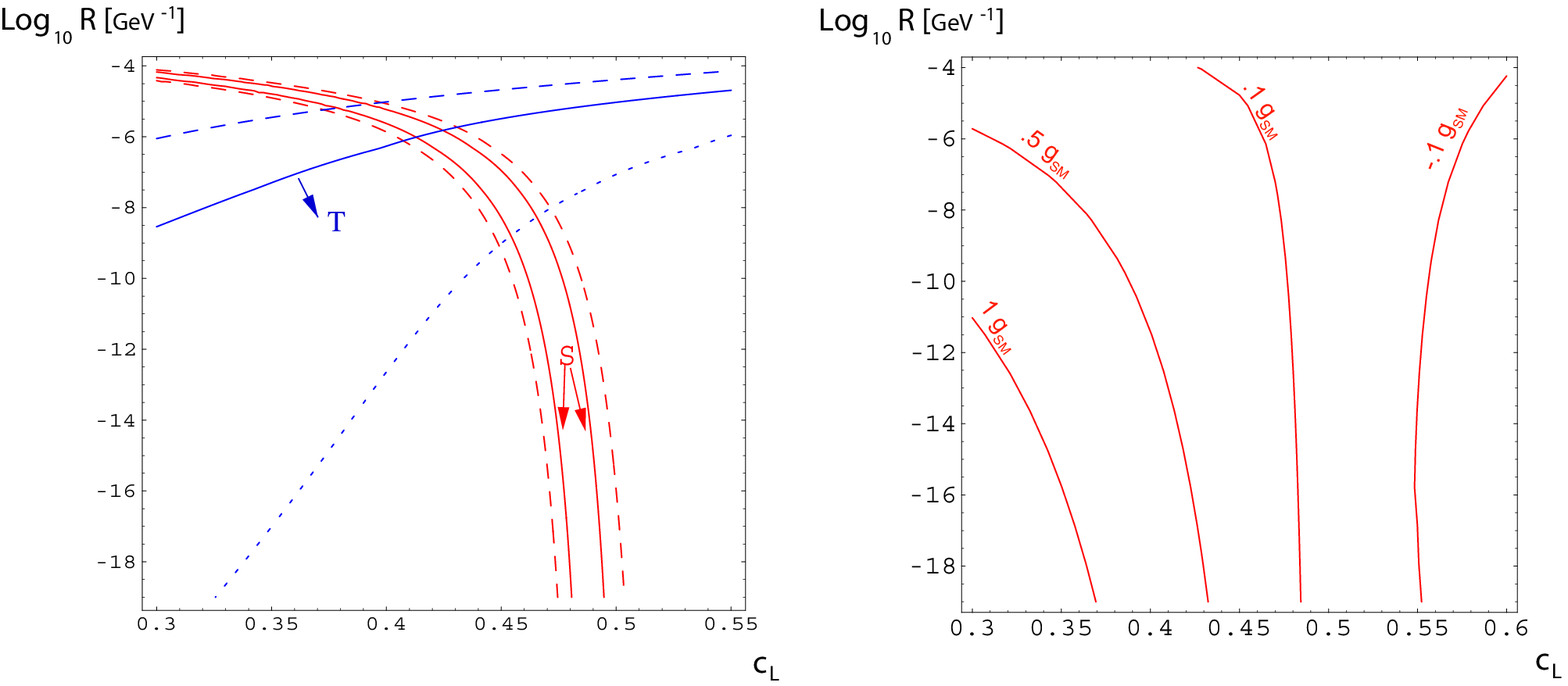}
\end{center}
\caption{
On the left, contours of $S$ (red), for $|S|=0.25$ (solid) and $0.5$
(dashed) and $T$ (blue), for  $|T|=0.1$ (dotted), $0.3$ (solid) and
$0.5$ (dashed), as function of $c_L$ and $R$.
On the right, contours for the generic suppression of fermion couplings
to the first resonance with respect to the SM value. In particular we
plotted the couplings of a lh down--type massless quark with the $Z'$.
The region for $c_L$, allowed by $S$, is between $0.43\div 0.5$, where
the couplings are suppressed at least by a factor of 10.}
\label{fig:coupR}
\end{figure}

In Fig.~\ref{fig:NDA} we have plotted the value of the NDA scale
(\ref{eq:NDA}) as well as the mass of the first resonance in the
$(c_L-R)$ plane.
Increasing $R$ also affects the oblique corrections.
However, while it is always possible to reduce $S$ by delocalizing the
fermions, $T$ increases and puts a limit on how far $R$ can be raised.
One can also see form Fig.~\ref{fig:coupR} that in the region where
$|S|<0.25$, the coupling of the first resonance with the light fermions
is generically suppressed to less than $10\%$ of the SM value.
This means that the LEP bound of $2$~TeV for SM--like $Z^\prime$ is
also decreased by a factor of 10 at least (the correction to the
differential cross section is roughly proportional to $g^2/M_{Z'}^2$).
In the end, values of $R$ as large as $10^{-7}$~GeV$^{-1}$ are allowed,
where the resonance masses are around $600$~GeV.
So, even if, following the analysis of~\cite{Papucci}, we take into
account a factor of roughly $1/4$ in the NDA scale, we see that the
appearance of strong coupling regime can be delayed up to $10$~TeV.  At
the LHC it will be very difficult to probe $WW$ scattering above 3 TeV.

\section{Flavor issues}

Finally we will address some issues about flavor
physics arising in this scenario.
First we will consider 
the eventual presence of flavor changing neutral currents
(FCNC) induced either by higher dimension operators or by
non--universal corrections.
Next, we will briefly discuss the problems surrounding the inclusion of
the third family of quarks in the picture. 

A generic 4-fermi operator would be given by the following expression
in this model:
\beq
\int d^4x\ \frac{1}{\Lambda_{FCNC}^2} \ \bar{\chi}_{u_L} \gamma^\mu
\chi_{u_L}
\ \bar{\chi}_{d_L} \gamma_\mu \chi_{d_L}.
\eeq
After putting in the wave functions of the zero modes one can estimate 
the scale of suppression of the flavor changing operators as a 
function of the $c$'s:
\beq
\Lambda_{FCNC}^2 \approx \frac{(R/R')^{2-4c_L} \log (R'/R)}{R'^2}.
\eeq
For $1/R'\sim 1$~TeV, to get a suppression factor of $10^3$~TeV, $c_L$
would have to be bigger than $0.57$.
Clearly the values of $c_L$ used to reduce the $S$ parameter do not
fulfill this criterion, which means that the set-up fails to naturally
explain the absence of FCNC and additional flavor symmetries in 5D
would be necessary. It is however relatively easy to impose such a
flavor symmetry in the bulk and on the TeV brane
and naturally break it close to the Planck brane. Due
to the small overlap of the fermion wavefunctions on the Planck brane,
the suppression scale of the four-Fermi operators will be significantly
increased.

The major challenge facing Higgsless models is the incorporation of the
third family of quarks.
There is a tension~\cite{BN,ADMS} in obtaining a large top quark mass
without
deviating from the observed bottom couplings with the $Z$. It can be
seen in the following way. The top
quark mass is proportional both to the Dirac mixing $M_D$ on the TeV
brane
and the overall scale of the extra dimension set by $1/R'$. For
$c_L\sim 0.5$ (or larger) it is in fact impossible to obtain a heavy
enough top
quark
mass (at least for $g_{5R}=g_{5L}$). The reason is that for $M_DR' \gg
1$
the light mode mass saturates at
\begin{equation}
m_{top}^2 \sim \frac{2}{R'^2 \log \frac{R'}{R}}\,,
\end{equation}
which gives for this case $m_{top}\leq \sqrt{2} M_W$. Thus one needs to
localize the top and the bottom quarks closer to the TeV brane.
However, even
in this case a sizable Dirac mass term on the TeV brane is needed to
obtain a heavy enough top quark. The consequence of this mass term is
the
boundary condition for the bottom quarks
\begin{equation}
\chi_{bR}= M_D R'\, \chi_{bL}.
\end{equation}
This implies that if $M_D R' \sim 1$ then the left handed bottom quark
has a sizable component also living in an $SU(2)_R$ multiplet, which
however
has a coupling to the $Z$ that is different from the SM value. Thus
there
will be a large deviation in the $Zb_L\bar{b}_L$. Note, that the same
deviation will not appear in the $Zb_R\bar{b}_R$ coupling, since
the extra kinetic term introduced on the Planck brane to split  top and
bottom will imply that the right handed $b$ lives mostly in the induced
fermion on the Planck brane which has the correct coupling to the $Z$.

The only way of
getting around this problem would be to raise the value of $1/R'$, and
thus
lower the necessary mixing on the TeV brane needed to obtain a heavy
top quark. One way of raising the value of $1/R'$ is by increasing
the ratio $g_{5R}/g_{5L}$ (at the price of making also the gauge KK
modes
heavier and thus the theory more strongly coupled).

Another generic problem arising from the large value of the
top-quark mass in models with warped extra dimensions
comes from the isospin violations in the KK sector of the top and the
bottom quarks. If the spectrum of the
top and bottom KK modes is not sufficiently degenerate, the loop
corrections involving these KK modes
to the $T$-parameter could be large.  This possibility was first pointed out
in~\cite{ADMS,ACP}.

\section{Conclusions}

We have discussed the possibility of breaking the electroweak symmetry 
by boundary conditions rather than with a scalar Higgs. We have found that
this may indeed be possible in theories with extra dimensions, and that in 
this case the scale of unitarity violation due to the absence of the
Higgs scalar could be significantly delayed due to the exchange of 
the massive KK modes of the gauge bosons. In order to find a realistic
model the presence of the custodial $SU(2)$ symmetry has to be guaranteed,
which leads us naturally to the warped higgsless model. In this model
the W/Z mass ratio is automatically the correct one, and there is a simple
way to introduce a splitting between the lightest KK modes (to be identified
with the ordinary W and Z) and the next ones. We have shown that fermions
can be similarly incorporated into this picture. In the simplest model
the prediction for the electroweak $S$-parameter comes out to be too large,
however one can modify the structure of the light fermions to be able
to suppress the $S$-parameter. The main unresolved issue is the 
incorporation of a sufficiently heavy top quark without inducing a large
shift in the $Zb\bar{b}$ coupling.

\section{Acknowledgements}

I thank Giacomo Cacciapaglia, Christophe Grojean,
Jay Hubisz, Hitoshi Murayama, Luigi Pilo, Yuri Shirman and John Terning
for collaborations on~\cite{CGMPT,CGPT,CGHST,CCGT,CCGT2} which 
were summarized in this talk. I thank the organizers of the 
SUSY 2004 conference in Tsukuba, Japan for inviting me and for
providing a stimulating athmosphere.
This research is supported in part by the DOE OJI grant DE-FG02-01ER41206 and in part
by the NSF grants PHY-0139738 and PHY-0098631.

\bibliographystyle{plain}

\end{document}